\documentclass[pra,preprint,amsmath,amssymb]{revtex4}
\usepackage{graphicx}
\usepackage{bm}
\usepackage{amsmath,amssymb}
\usepackage{color}

\begin{document}

\title{Continuous-variable blind quantum computation} 

\author{Tomoyuki Morimae}
\email{morimae@gmail.com}
\affiliation
{Department of Physics,
Imperial College London,
London SW7 2BW, United Kingdom}
\date{\today}
            
\begin{abstract}
Blind quantum computation is a secure delegated
quantum computing protocol where 
Alice who does not have sufficient quantum technology
at her disposal
delegates her computation to Bob who has a fully-fledged quantum
computer in such a way that Bob cannot learn anything about
Alice's input, output, and algorithm.
Protocols of blind quantum computation
have been proposed for several qubit measurement-based computation models,
such as the graph state model, the Affleck-Kennedy-Lieb-Tasaki model,
and the Raussendorf-Harrington-Goyal topological model.
Here, we consider blind quantum computation
for the continuous-variable measurement-based model.
We show that blind quantum computation is possible for
the infinite squeezing case. We also show 
that the finite squeezing causes no additional problem in the blind setup
apart from the one inherent to the continuous-variable
measurement-based quantum computation.
\end{abstract}
\maketitle  

\section{Introduction}
When scalable quantum computers are realized, 
they will be used in ``cloud computing style"
since only limited number of people will be able
to possess quantum computers.
Blind quantum computation~\cite{Childs,Arrighi,blindcluster,Aha,
TVE,Vedran,Barz,blind_Raussendorf,FK,measuringAlice}
provides a solution to the issue of the client's security in such a
cloud quantum computation.
Blind quantum computation is a new secure protocol
which enables Alice who does not have enough quantum technology
to delegate her computation to Bob who has a fully-fledged
quantum computer in such a way that Bob cannot learn anything
about Alice's input, output,
and algorithm.
A protocol of the 
unconditionally secure universal blind quantum computation for
almost classical Alice was first
proposed in Ref.~\cite{blindcluster}
by using the measurement-based quantum 
computation (MBQC) on the cluster state~\cite{cluster,cluster2,Raussendorf_PhD},
and later generalized to other resource states
such as the Affleck-Kennedy-Lieb-Tasaki state~\cite{TVE,MiyakeAKLT,AKLT}
and the three-dimensional Raussendorf-Harrington-Goyal 
state~\cite{Raussendorf_PRL,Raussendorf_NJP,Raussendorf_Ann,Li_finiteT,
FM_finiteT}
which enables the topological protection~\cite{blind_Raussendorf,FK}.
A subroutine which eases Alice's burden was invented~\cite{Vedran}. 
Also, a verification scheme which tests 
Bob's honesty was proposed~\cite{FK}. 
The proof-of-principle experiment of the original protocol~\cite{blindcluster}
was realized
in an optical system~\cite{Barz}.

In this paper, we consider the continuous-variable (CV) version
of the blind
quantum computation.
The CV cluster MBQC
was proposed in Refs.~\cite{canonical_full,canonical_letter,review}.
There, 
\begin{eqnarray*}
|+\rangle\equiv\frac{1}{\sqrt{2}}(|0\rangle+|1\rangle)
\end{eqnarray*}
state of a single qubit is replaced with the 
zero momentum state $|0\rangle_p$
of a single mode (qumode),
and the two-mode gate $e^{iq\otimes q}$ plays the role of
the qubit Controlled-Z gate,  
\begin{eqnarray*}
|0\rangle\langle0|\otimes I+|1\rangle\langle1|\otimes Z.
\end{eqnarray*}
Experimental demonstrations of
building blocks of
the CV cluster MBQC were already 
achieved~\cite{experiment1,experiment2,experiment3,
experiment5,experiment6}.

We show that blind quantum computation is possible in the infinite
squeezing case. 
We also consider the finite squeezing case, and show that
the finite squeezing causes no problem 
apart from the additional errors which come from the redundancy
of gates required for the blindness.
Since these errors are those
even the non-blind CV MBQC
has to cope with for its scalability,
we conclude that
the finite squeezing does not cause any fundamental problem
in principle.

This paper is organized as follows.
In the next section we will briefly review the CV cluster MBQC.
We also review the qubit blind quantum computation in Sec.~\ref{sec_blind}.
Then we explain our protocol in Sec.~\ref{sec_protocol},
and show its correctness (Sec.~\ref{sec_correctness}) 
and blindness (Sec.~\ref{sec_blindness}).
Discussions are given in Sec.~\ref{sec_discussion}.

\section{CV cluster MBQC}
Let us briefly review
the CV cluster MBQC
proposed in Refs.~\cite{canonical_letter,canonical_full}.
Let $q$ and $p$ be the quadrature ``position" and ``momentum"
operators, respectively, satisfying 
the canonical commutation relation
\begin{eqnarray*}
[q,p]=i.
\end{eqnarray*}
We also define the Weyl-Heisenberg operators
\begin{eqnarray*}
X(s)&\equiv&\exp[-isp],\\
Z(s)&\equiv&\exp[isq],
\end{eqnarray*}
with $s\in {\mathbb R}$,
where
\begin{eqnarray*}
X(s)|t\rangle_q&=&|t+s\rangle_q,\\
Z(s)|t\rangle_p&=&|t+s\rangle_p.
\end{eqnarray*}
Here, $|t\rangle_q$ and $|t\rangle_p$
are eigenvectors of $q$ and $p$ with the eigenvalue $t$,
respectively.
They satisfy
\begin{eqnarray*}
X(s)Z(t)=e^{-ist}Z(t)X(s),\\
qX(s)=X(s)(q+s),\\
pZ(s)=Z(s)(p+s).
\end{eqnarray*}
These Weyl-Heisenberg operators are CV analog of the qubit
Pauli operators.
The Fourier transform operator 
$F$
is defined by
\begin{eqnarray*}
F\equiv\exp\Big[i(q^2+p^2)\frac{\pi}{4}\Big],
\end{eqnarray*}
with 
\begin{eqnarray*}
F|s\rangle_q=|s\rangle_p.
\end{eqnarray*}
This operator is the CV analog of the qubit Hadamard operator.
However, special cares are needed because
$F$ is not Hermitian and
\begin{eqnarray*}
F^2|s\rangle_q&=&|-s\rangle_q,\\
F^2|s\rangle_p&=&|-s\rangle_p,\\
F^4&=&I,\\
F^\dagger q F&=&-p,\\
F^\dagger p F&=&q,\\
Z(m)F&=&FX(m),\\
X(m)F&=&FZ(-m).
\end{eqnarray*}
The CV version of the Controlled-$Z$ gate
are defined by
\begin{eqnarray*}
CZ\equiv\exp(iq\otimes q).
\end{eqnarray*}
Note that we use the symbol $CZ$
for both qubit $CZ$ and CV $CZ$.
But no confusion will occur because they can be distinguished
from the context.
The CV version of the Controlled-$X$ gate
is defined by
\begin{eqnarray*}
CX\equiv\exp(-iq\otimes p).
\end{eqnarray*}

\begin{figure}[htbp]
\begin{center}
\includegraphics[width=0.4\textwidth]{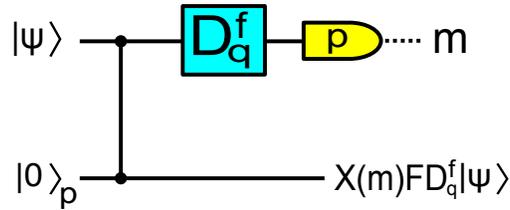}
\end{center}
\caption{ 
(Color online.)
The CV teleportation gate.
} 
\label{teleportation}
\end{figure}

The elementary block of the CV cluster MBQC is the
teleportation gate given in Fig.~\ref{teleportation}.
Here, 
\begin{eqnarray*}
D_q^f\equiv\exp[if(q)],
\end{eqnarray*}
and $f$ is a polynomial of $q$.
Note that $D_q^f$ and $D_p^f$
are obtained from $FD_q^f$, since
\begin{eqnarray*}
(FD_q^0)^3FD_q^f&=&D_q^f,\\
(FD_q^0)^2(FD_{-q}^{f})(FD_q^0)&=&D_p^f.
\end{eqnarray*}
Furthermore,
$e^{isq^k/k}$ $(k=1,2,3)$ and 
$e^{isp^k/k}$ $(k=1,2,3)$ are single-mode universal~\cite{Lloyd}.
Hence
\begin{eqnarray*}
R_q(v)\equiv F\exp\Big[i\Big(aq+b\frac{q^2}{2}+c\frac{q^3}{3}\Big)\Big]
\end{eqnarray*}
is single-mode universal,
where $v=(a,b,c)$.
Addition of $CZ$ enables all multi-mode universality.

\begin{figure}[htbp]
\begin{center}
\includegraphics[width=0.45\textwidth]{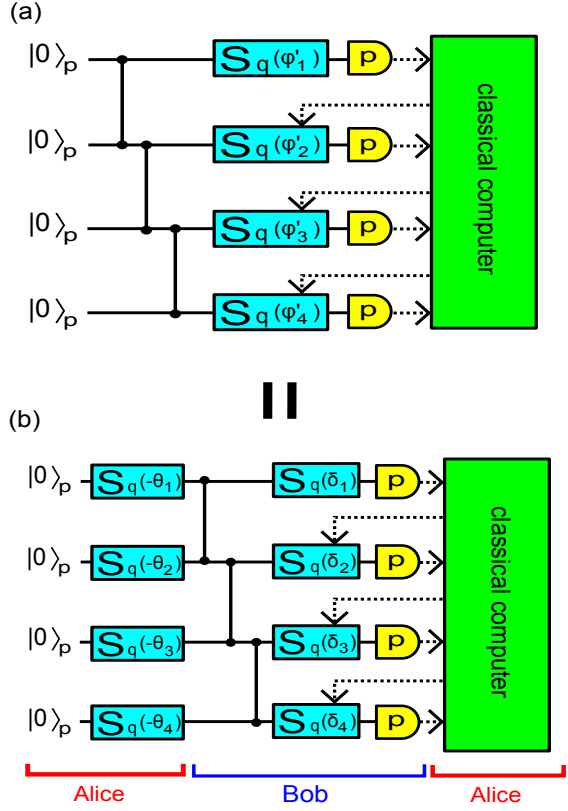}
\end{center}
\caption{(Color online.) 
(a): $S_q(\phi)$ stands for $F^\dagger R_q(\phi)$.
This circuit is universal if $\{\phi'_j\}$ are appropriately chosen according
to the previous measurement results.
(b): The blind version of (a). Obviously, $S_q$'s commute with $CZ$'s,
and therefore (b) is equivalent to (a).
} 
\label{blind}
\end{figure}

Let us explain how to compensate the byproduct error $X(m)$.
Note that
\begin{eqnarray*}
R_q(v)X(m)&=&Z(m)R_{q+m}(v),\\
&=&Z(m)R_q(M_mv),
\end{eqnarray*}
where
\begin{eqnarray*}
M_m=
\left(
\begin{array}{ccc}
1&m&m^2\\
0&1&2m\\
0&0&1
\end{array}
\right)
\end{eqnarray*}
and its inverse is
\begin{eqnarray*}
M_m^{-1}=
\left(
\begin{array}{ccc}
1&-m&m^2\\
0&1&-2m\\
0&0&1
\end{array}
\right).
\end{eqnarray*}
Therefore, if we want to implement
$R_q(v)$, and if there is the byproduct
$X(m)$, we have only to implement
$R_q(M_m^{-1}v)$.
Furthermore, we can show
\begin{eqnarray*}
CZ(X(m)\otimes I)&=&(X(m)\otimes Z(m))CZ,\\
CZ(I\otimes X(m))&=&(Z(m)\otimes X(m))CZ.
\end{eqnarray*}
Therefore, the byproducts can be sent forward through
$CZ$ gates.
In short, Fig.~\ref{blind} (a) is universal
if the feed-forwarding is appropriately done.

The application of $D_q^f$ followed by the measurement of $p$
is equivalent to the measurement of the
observable $(D_q^f)^\dagger p D_q^f$.
Therefore, to implement the gate $e^{isq}$ in Fig.~\ref{teleportation}, 
we measure
\begin{eqnarray*}
e^{-isq}pe^{isq}=p+s.
\end{eqnarray*}
It can be measured easily with a homodyne detection.
To implement the gate $e^{isq^2/2}$
in Fig.~\ref{teleportation},
we measure
\begin{eqnarray*}
e^{-isq^2/2}pe^{isq^2/2}=p+sq.
\end{eqnarray*}
It can also be measured with
a homodyne detection in a rotated quadrature basis.
In principle, the gate $e^{isq^3/3}$
can be implemented
in Fig.~\ref{teleportation}
by measuring
\begin{eqnarray*}
e^{-isq^3/3}pe^{isq^3/3}=p+sq^2.
\end{eqnarray*}

Finally, let us notice that
the zero-momentum state $|0\rangle_p$ is not realistic,
and normally $|0\rangle_p$ is approximated by the 
finitely squeezed vacuum state
\begin{eqnarray*}
|0,\Omega\rangle_p=\frac{1}{(\pi\Omega^2)^{1/4}}
\int dp~ e^{-\frac{p^2}{2\Omega^2}}|p\rangle_p.
\end{eqnarray*}
This finite squeezing causes errors in the CV cluster 
MBQC~\cite{canonical_letter,canonical_full}.

\section{Blind quantum computation}
\label{sec_blind}
Let us also briefly review the basic idea of the
original blind quantum computation protocol
of Ref.~\cite{blindcluster}.
For details, see Refs.~\cite{blindcluster,
TVE,Vedran,Barz,blind_Raussendorf,FK,measuringAlice}.
Alice, the client, has a quantum device which emits  
randomly-rotated single qubit states and a classical computer.
Bob, the server, has a full quantum power.
Let us assume that Alice wants to perform
the cluster MBQC on the $N$-qubit graph state $|G\rangle$
with measurement angles $\{\phi_j\}_{j=1}^N$.
If Alice sends Bob $\{\phi_j\}_{j=1}^N$, and Bob creates $|G\rangle$,
the delegated quantum computation is of course possible.
However, obviously, in this case Bob can learn Alice's privacy.
Hence they run the following protocol:
\begin{enumerate}
\item
Alice sends Bob $N$ randomly-rotated single-qubit states
$\{|+_{\theta_j}\rangle\}_{j=1}^N$,
where
\begin{eqnarray*}
|+_\theta\rangle\equiv\frac{1}{\sqrt{2}}(|0\rangle+e^{i\theta}|1\rangle)
=e^{-iZ\theta/2}|+\rangle
\end{eqnarray*}
and 
\begin{eqnarray*}
\theta_j\in\Big\{\frac{k\pi}{4}~\Big|k=0,1,...,7\Big\}
\end{eqnarray*}
is a random angle
which is hidden to Bob.
\item
Bob applies $CZ$ gates among them. Since $CZ$ commutes
with $e^{-iZ\theta/2}$, what Bob obtains is
\begin{eqnarray*}
&&\Big(\bigotimes_{e\in E}CZ_e\Big)
\Big(\bigotimes_{j=1}^Ne^{-iZ_j\theta_j/2}\Big)
|+\rangle^{\otimes N}\\
&=&
\Big(\bigotimes_{j=1}^Ne^{-iZ_j\theta_j/2}\Big)
\Big(\bigotimes_{e\in E}CZ_e\Big)
|+\rangle^{\otimes N}\\
&=&
\Big(\bigotimes_{j=1}^Ne^{-iZ_j\theta_j/2}\Big)
|G\rangle,
\end{eqnarray*}
where $E$ is the set of edges of $G$,
and the subscript $j$ of $Z_j$ means the operator acts on $j$th qubit.
\item
For $j=1$ to $N$ in turn:
\begin{enumerate}
\item
Alice sends Bob 
\begin{eqnarray*}
\delta_j\equiv\theta_j+\phi_j'+r_j\pi,
\end{eqnarray*}
where $\phi_j'$ is the modification of $\phi_j$
which includes appropriate feedforwardings
(byproduct corrections)
and $r_j\in\{0,1\}$
is a random binary.
\item
Bob does the measurement in the $\{|\pm_{\delta_j}\rangle\}$ basis,
and returns the measurement result to Alice.
\end{enumerate}
\end{enumerate}


It was shown in Ref.~\cite{blindcluster} that 
this protocol is correct. 
Here, correct means that if Bob is honest
Alice obtains the correct outcome.
In fact, if Bob measures $k$th qubit in the
$\{|\pm_{\delta_k}\rangle\}$ basis,
\begin{eqnarray*}
&&\langle \pm_{\delta_k}|
\Big(\bigotimes_{j=1}^Ne^{-iZ_j\theta_j/2}\Big)|G\rangle\\
&=&
\langle \pm|e^{iZ_k\delta_k/2}
\Big(\bigotimes_{j=1}^Ne^{-iZ_j\theta_j/2}\Big)|G\rangle\\
&=&
\Big(\bigotimes_{j\neq k}
e^{-iZ_j\theta_j/2}\Big)
\langle \pm|e^{iZ_k\delta_k/2}
e^{-iZ_k\theta_k/2}
|G\rangle\\
&=&
\Big(\bigotimes_{j\neq k}
e^{-iZ_j\theta_j/2}\Big)
\langle \pm|e^{iZ_k(\phi_k'+r_k\pi)/2}
|G\rangle\\
&=&
\Big(\bigotimes_{j\neq k}
e^{-iZ_j\theta_j/2}\Big)
\langle \pm_{\phi_k'}|Z^{r_k}_k|
G\rangle,
\end{eqnarray*}
which means that Bob effectively does the
$\{|\pm_{\phi_k'}\rangle\}$
basis measurement
with the error $Z^{r_k}$.
The error $Z^{r_k}$ just flips the bit of the measurement result,
and therefore it can be compensated later.

It was also shown that
the protocol is blind~\cite{blindcluster}. 
Here, blind intuitively means that whatever Bob does,
Bob cannot learn anything about Alice's input, output,
and algorithm.
Intuitive proof of the blindness is as follows:
What Bob obtains are quantum states 
$\{|+_{\theta_j}\rangle\}_{j=1}^N$
and classical messages
$\{\delta_j\}_{j=1}^N$.
Hence Bob's state is
\begin{eqnarray*}
\sum_{\phi_1,...,\phi_N}
\sum_{r_1,...,r_N}
\bigotimes_{j=1}^N
|+_{\phi_j'+r_j\pi}\rangle\langle+_{\phi_j'+r_j\pi}|
=I^{\otimes N},
\end{eqnarray*}
which means that Bob cannot learn anything about $\{\phi_j\}_{j=1}^N$
whichever POVM he does on his system.

\begin{figure}[htbp]
\begin{center}
\includegraphics[width=0.4\textwidth]{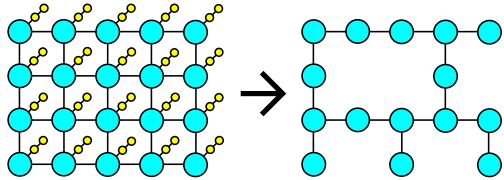}
\end{center}
\caption{(Color online.) 
The hair implantation technique~\cite{blind_Raussendorf}.
Left: A two-qubit graph state (``hair") indicated by yellow
is attached to each blue qubit of the square
graph state.
Right: A desired graph state can be carved out from
the blue square graph.
} 
\label{hair}
\end{figure}

\begin{figure}[htbp]
\begin{center}
\includegraphics[width=0.25\textwidth]{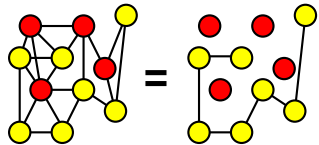}
\end{center}
\caption{(Color online.) 
The graph hiding technique~\cite{FK}.
Yellow qubits are $|+_\theta\rangle$, whereas
red qubits are $|0\rangle$ or $|1\rangle$.
Bob applies $CZ$ gates on all edges of the left graph,
but actually he obtains the right graph state,
and he does not know its geometry.
} 
\label{hiding}
\end{figure}

In order to guarantee Alice's privacy, 
the geometry of the graph $G$ must be secret to Bob.
There are three ways of doing it.
First one is to use the brickwork state~\cite{blindcluster}.
It is a certain two-dimensional graph state which is universal
with only $\{|\pm_\theta\rangle\}$ basis measurements
for $\theta\in\{\frac{k\pi}{4}|k=0,1,...,7\}$.
Second one is to implant a ``hair" to each qubit of the 
regular lattice graph state~\cite{blind_Raussendorf}.
For example, let us consider the left graph state of Fig.~\ref{hair}.
We can simulate $Z$ measurement and any $X-Y$ plane measurement
on any blue qubit with only $X-Y$ plane measurements
on yellow and blue qubits.
Hence we can ``carve out" a specific graph state from
the square lattice of blue qubits as is shown in the right
of Fig.~\ref{hair}.
Third one is
so called ``the graph hiding technique"~\cite{FK}.
By using this technique,
Alice can have Bob prepare any graph state in such a way
that Bob cannot learn the geometry of the graph.
This technique is based on the simple idea that $CZ$ does not
create entanglement if one of the qubits is $|0\rangle$ or $|1\rangle$:
\begin{eqnarray*}
CZ(|\psi\rangle\otimes|+\rangle)&=&
CZ(|\psi\rangle\otimes|+\rangle),\\
CZ(|\psi\rangle\otimes|-\rangle)&=&
(I\otimes Z)CZ(|\psi\rangle\otimes|-\rangle),\\
CZ(|\psi\rangle\otimes|0\rangle)&=&|\psi\rangle\otimes|0\rangle,\\
CZ(|\psi\rangle\otimes|1\rangle)&=&Z|\psi\rangle\otimes|1\rangle.
\end{eqnarray*}
Therefore, if Alice hides several qubits in $|0\rangle$ or $|1\rangle$
into the set of qubits she initially sends to Bob,
she can let Bob create her desired graph state.
Since Bob cannot distinguish $|0\rangle$, $|1\rangle$,
and eight $|+_\theta\rangle$ states, Bob cannot know when
he entangles qubits (Fig.~\ref{hiding}). 

\section{CV blind protocol}
\label{sec_protocol}
Now let us consider
the CV blind protocol.
We here describe the ideal version and later
consider realistic situations.
Our protocol runs as follows:
\begin{enumerate}
\item
Alice sends Bob 
\begin{eqnarray*}
\Big\{
S_q(-\theta_j)|0\rangle_p
\Big\}_{j=1}^N,
\end{eqnarray*}
where 
$\theta_j=(a_j,b_j,c_j)$
is randomly chosen from ${\mathbb R}^3$
and
$S_q(v)=F^\dagger R_q(v)$.
\item
Bob applies $CZ$ gates.
\item
Alice and Bob might choose ``the brickwork", ``the hair implantation technique",
or ``the graph hiding technique".
Irrespective of their choice, we can assume without loss of generality
that Bob has the ``encrypted" CV graph state
\begin{eqnarray*}
\Big[
\bigotimes_{j=1}^N
X^j(\xi_j)
Z^j(\eta_j)
S_q^j(-\theta_j)
\Big]
|G\rangle,
\end{eqnarray*}
where $|G\rangle$ is the $N$-qumode CV graph state,
and the subscript $j$ of $X^j$
means it acts on the $j$th qumode.
\item
For $j=1$ to $N$ in turn:
\begin{enumerate}
\item
Let $\phi_j\equiv(\alpha_j,\beta_j,\gamma_j)$ be 
Alice's computational parameters,
and
Let $\phi_j'\equiv(\alpha'_j,\beta_j',\gamma_j')$ be 
the one including feedforwardings.
Alice sends Bob
$\delta_j=M_{\xi_j}^{-1}w_j$
where 
\begin{eqnarray*}
w_j=
\left(
\begin{array}{l}
\alpha_j'+a_j-\eta_j+r_j\\
\beta_j'+b_j\\
\gamma_j'+c_j
\end{array}
\right)
\end{eqnarray*}
and $r_j\in{\mathbb R}$ is a random real number.
\item
Bob applies $S_q(\delta_j)$ on $j$th qumode
and does the $p$ measurement on it.  
(Or he directly measures 
$S_q^\dagger(\delta_j)pS_q(\delta_j)$ 
of the $j$th qumode.)
He sends the measurement result to Alice.
\end{enumerate}
\end{enumerate}

\section{Correctness}
\label{sec_correctness}
Let us show the correctness of our protocol.
See Fig.~\ref{blind} (b),
which is the circuit representation of our protocol.
Since $S_q$ commutes with $CZ$, Fig.~\ref{blind} (b) is equivalent to 
Fig.~\ref{blind} (a).
Note that the equivalence between (a) and (b) in Fig.~\ref{blind}
is based on only the commutativity between $S_q$ and $CZ$, 
and therefore it holds even if we replace each input $|0\rangle_p$
with its finitely squeezed version.
Hence, the finite squeezing does not cause any additional
effect here.

More precisely, note that the following is true
for any state $|\psi\rangle$:
\begin{eqnarray*}
&&{}_p\langle p|
S_q^j(\delta_j)X^j(\xi_j)
Z^j(\eta_j)S_q^j(-\theta_j)|\psi\rangle\\
&=&
{}_p\langle p|
X^j(\xi_j)
Z^j(\eta_j)
S_q^j(M_{\xi_j}M_{\xi_j}^{-1}w_j)
S_q^j(-\theta_j)|\psi\rangle\\
&=&
{}_p\langle p|
X^j(\xi_j)
Z^j(\eta_j)
S_q^j(w_j)
S_q^j(-\theta_j)|\psi\rangle\\
&=&
{}_p\langle p|
\exp\Big[
i\Big\{\\
&&(\alpha_j'+a_j-\eta_j+r_j)q
+(\beta_j'+b_j)\frac{q^2}{2}
+(\gamma_j'+c_j)\frac{q^3}{3}\\
&&+\eta_jq
-a_jq
-b_j\frac{q^2}{2}
-c_j\frac{q^3}{3}
\Big\}
\Big]|\psi\rangle\\
&=&
{}_p\langle p|
\exp[ir_jq]
\exp\Big[
i\Big\{
\alpha_j'q
+\beta_j'\frac{q^2}{2}
+\gamma_j'\frac{q^3}{3}
\Big\}
\Big]|\psi\rangle\\
&=&
{}_p\langle p|
\exp[ir_jq]
S_q^j(\phi_j')|\psi\rangle\\
&=&
{}_p\langle p-r_j|
S_q^j(\phi_j')|\psi\rangle.
\end{eqnarray*}
Hence, Bob effectively does the correct MBQC
except for the fact that if the measurement
result is $p$, the byproduct which comes from this measurement
is not $X(p)$ but $X(p-r_j)$,
which can be compensated by changing the following measurement parameters.
Since the above equation is true for any state $|\psi\rangle$,
the situation does not change even if the squeezing is finite.

\begin{figure}[htbp]
\begin{center}
\includegraphics[width=0.4\textwidth]{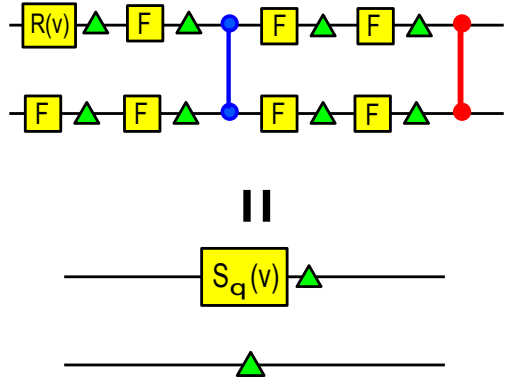}
\end{center}
\caption{(Color online.) 
The implementation of $S_q(v)\otimes I$.
Blue two-qubit gate is $CZ$.
Red two-qubit gate is $CZ^\dagger$.
Green triangles are byproducts.
} 
\label{brick}
\end{figure}

\begin{figure}[htbp]
\begin{center}
\includegraphics[width=0.4\textwidth]{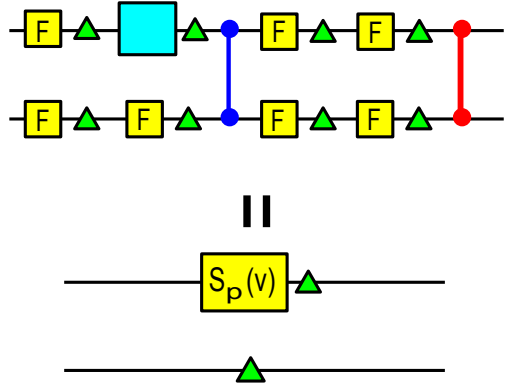}
\end{center}
\caption{(Color online.) 
The implementation of $S_p(v)\otimes I$.
The blue box means $R_{-q}(M_m^{-1}v)$.
} 
\label{brick2}
\end{figure}

\begin{figure}[htbp]
\begin{center}
\includegraphics[width=0.4\textwidth]{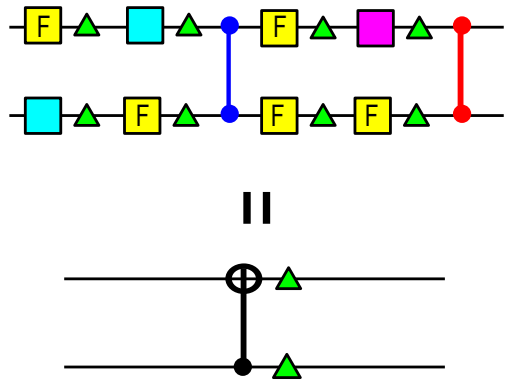}
\end{center}
\caption{(Color online.) 
The implementation of $CX$.
The blue boxes are $Fe^{iq^2/2}$ up to byproduct corrections.
The purple box is $Fe^{-iq^2/2}$ up to byproduct corrections.
} 
\label{brick3}
\end{figure}

The brickwork implementation for the CV blind protocol
is shown in Fig.~\ref{brick}, \ref{brick2}, and \ref{brick3}.
Since $CZ\cdot CZ \neq I$ for the CV case,
we cannot directly generalize the qubit brickwork state of Ref.~\cite{blindcluster}.
In particular, we need $CZ$ and $CZ^\dagger$
as is shown in Figs.~\ref{brick}, \ref{brick2}, and \ref{brick3}.

The hair implantation technique also works
if we implant four-qumode hair on each qumode,
since the measurement of $q$ on a qumode in a CV graph state
removes that qumode~\cite{canonical_full},
and a $q$ measurement can be simulated only with
$S_q^\dagger p S_q$ measurements by using the following relations:
\begin{eqnarray*}
F\cdot Fe^{iq^2/2}\cdot Fe^{iq^2/2}\cdot F&=&e^{iq^2/2}e^{ip^2/2},\\
e^{-ip^2/2}e^{-iq^2/2}pe^{iq^2/2}e^{ip^2/2}&=&q.
\end{eqnarray*}

The graph hiding technique for qubits
can also be generalized to CV,
since
\begin{eqnarray}
CZ(|\psi\rangle\otimes |s\rangle_p)&=&
(I\otimes Z(s))CZ(|\psi\rangle\otimes|0\rangle_p),\nonumber\\
CZ(|\psi\rangle\otimes |s\rangle_q)&=&
(Z(s)|\psi\rangle)\otimes|s\rangle_q.
\label{cut}
\end{eqnarray}
Therefore, Alice can have Bob create
a graph state where $Z(s)$
are applied on some qumodes
in such a way that Bob cannot know the graph geometry.

Finally, let us consider the effect of
the finite squeezing.
As we have seen, the Alice's prerotation technique
itself is valid for any
initial state (Fig.~\ref{blind}), and therefore the finite
squeezing does not cause any additional problem apart from
the original one inherent to
the non-blind CV MBQC~\cite{canonical_full,canonical_letter}.
If Alice and Bob choose the brickwork implementation
or the hair implantation technique,
again the finite squeezing does not cause any additional effect
since the brickwork blind quantum computation and the
hair implantation technique are nothing but
a normal cluster MBQC with some redundant gates.
(Of course, this redundancy accelerates the accumulation of errors, and therefore
requires more fault-tolerance,
but such a problem is not a specific problem to the blind CV MBQC.
Even the non-blind one ultimately needs enough
fault-tolerance for the scalability~\cite{canonical_full,canonical_letter,
Ohliger,experiment3}.)
Finally, regarding
the graph hiding technique, 
once the graph state is created, it is nothing but
a usual CV MBQC with errors.
If the squeezing is finite,
Eq.~(\ref{cut}) becomes not exact but
approximate one.
This causes additional errors on the created graph state,
but such errors are that even the non-blind CV MBQC can experience. 

\section{Blindness}
\label{sec_blindness}
What Bob obtains are
quantum states
$\{S_q(-\theta_j)|0\rangle_p\}_{j=1}^N$
and
classical messages
$\{\delta_j\}_{j=1}^N$.
Note that
\begin{eqnarray*}
\theta_j&=&M_{\xi_j}\delta_j-\phi_j'+\eta_je-r_je\\
&\equiv&k_j-r_je,
\end{eqnarray*}
where $e=(1,0,0)$.
Hence,
Bob's state is
\begin{eqnarray*}
&&\int
\prod_{j=1}^N dr_j
\bigotimes_{j=1}^N
S_q(-\theta_j)|0\rangle_p{}_p\langle 0|S_q^\dagger(-\theta_j)\\
&=&
\int
\prod_{j=1}^Ndr_j
\bigotimes_{j=1}^N
S_q(-k_j+r_je)|0\rangle_p{}_p\langle 0|S_q^\dagger(-k_j+r_je)\\
&=&
\int
\prod_{j=1}^Ndr_j
\bigotimes_{j=1}^N
S_q(-k_j)e^{ir_jq}|0\rangle_p{}_p\langle 0|e^{-ir_jq}S_q^\dagger(-k_j)\\
&=&
\int
\prod_{j=1}^Ndr_j
\bigotimes_{j=1}^N
S_q(-k_j)|r_j\rangle_p{}_p\langle r_j|S_q^\dagger(-k_j)\\
&=&
I^{\otimes N},
\end{eqnarray*}
which means that Bob's state is independent of $\{\phi_j\}_{j=1}^N$.

Note that the blindness holds also in the finite squeezed
case, since
\begin{eqnarray*}
&&\int
\prod_{j=1}^N dr_j
\bigotimes_{j=1}^N
S_q(-\theta_j)|0,\Omega_j\rangle_p{}_p\langle 0,\Omega_j|S_q^\dagger(-\theta_j)\\
&=&
\int
\prod_{j=1}^Ndr_j
\bigotimes_{j=1}^N
S_q(-k_j+r_je)|0,\Omega_j\rangle_p{}_p\langle 0,\Omega_j|S_q^\dagger(-k_j+r_je)\\
&=&
\int
\prod_{j=1}^Ndr_j
\bigotimes_{j=1}^N
S_q(-k_j)e^{ir_jq}|0,\Omega_j\rangle_p{}_p\langle 0,\Omega_j|e^{-ir_jq}S_q^\dagger(-k_j)\\
&=&
\int
\prod_{j=1}^Ndr_j
\bigotimes_{j=1}^N
S_q(-k_j)e^{ir_jq}T_{\Omega_j}|0\rangle_p{}_p\langle 0|T^\dagger_{\Omega_j}e^{-ir_jq}S_q^\dagger(-k_j)\\
&=&
\int
\prod_{j=1}^Ndr_j
\bigotimes_{j=1}^N
T_{\Omega_j}
\Big[
S_q(-k_j)e^{ir_jq}|0\rangle_p{}_p\langle 0|e^{-ir_jq}S_q^\dagger(-k_j)
\Big]
T^\dagger_{\Omega_j}.
\end{eqnarray*}
Here,
the operator
\begin{eqnarray*}
T_{\Omega}\equiv\frac{1}{(\pi\Omega^2)^{1/4}}
\int dt~~ e^{-\frac{t^2}{2\Omega^2}}e^{iqt}
\end{eqnarray*}
commutes with $S_q$.


\section{Discussion}
\label{sec_discussion}

\subsection{Implementation of $e^{iq^3/3}$}
In optical systems,
the implementation of $e^{iq^3/3}$
is much harder than those
of $e^{iq}$ and $e^{iq^2/2}$.
Hence it would be desirable for Alice to avoid
the implementation of $e^{iq^3/3}$ by herself.
There are two solutions.
One is that
Bob embeds many $e^{isq^3/3}|0\rangle_p$
with various $s$ into his resource state. 
If Alice uses the hair implantation technique or
the graph hiding technique,
Bob cannot know which $e^{isq^3/3}|0\rangle_p$
contributes to the computation.
The other is to use the relation
\begin{eqnarray}
Q^\dagger(t)e^{i\gamma q^3/3}Q(t)=e^{i\gamma'q^3/3},
\label{cubicle}
\end{eqnarray}
where 
\begin{eqnarray*}
Q(t)\equiv e^{-i\ln(t)(qp+pq)/2}
\end{eqnarray*}
is the squeezing and $t=(\gamma'/\gamma)^{1/3}$.
Since the squeezing
can be done blindly, Alice can have
Bob implement $e^{i\gamma'q^3/3}$ without allowing Bob to learn
$\gamma'$.

\subsection{Blind CV protocol for measuring Alice}
If the state measurement is relatively easy, we can consider
another blind quantum computation
protocol, where Bob creates the resource state
and Alice does the measurement~\cite{measuringAlice}.
One advantage of this protocol is that the security is guaranteed
by the no-signaling principle~\cite{nosignaling}, which is more fundamental than
quantum physics, and Alice does not need to verify her measurement
device (the device independence~\cite{deviceindependent}). 
The CV cluster MBQC is suitable for such a measuring Alice protocol,
since the measurements of
\begin{eqnarray*}
e^{-isq}pe^{isq}&=&p+s,\\
e^{-isq^2/2}pe^{isq^2/2}&=&p+sq
\end{eqnarray*}
are easily done with the homodyne detection.
The gate $e^{isq^3/3}$
can be implemented blindly by using
Eq.~(\ref{cubicle}).

\subsection{Temporal encoding}
\begin{figure}[htbp]
\begin{center}
\includegraphics[width=0.5\textwidth]{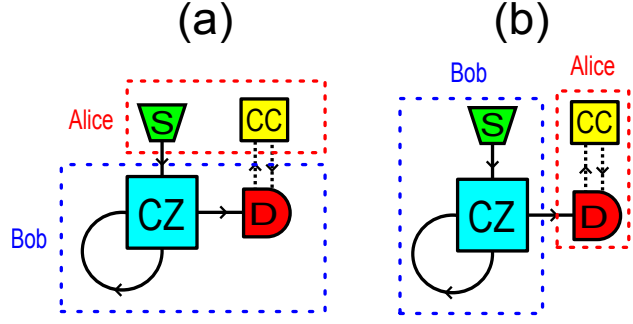}
\end{center}
\caption{(Color online.) 
Blind version of the temporal-encoding~\cite{temporal}.
S is the squeezed state source, D is the measurement device,
CZ is the machine which implements the $CZ$ gate,
and CC is a classical computer.
} 
\label{time}
\end{figure}

If we use the temporal degrees of freedom,
only a single $CZ$ machine is sufficient~\cite{temporal}.
As is shown in Fig.~\ref{time},
it is easy to see that
blind versions of such a temporal encoding implementation are
possible both for the preparing Alice (Fig.~\ref{time} (a)) 
and the measuring Alice (Fig.~\ref{time} (b)).

\acknowledgements
The author acknowledges JSPS for support.

\appendix
\section{Proof of Fig.~\ref{brick}}
\begin{eqnarray*}
&&
CZ^\dagger
\left(
\begin{array}{c}
PF\cdot PF\\
PF\cdot PF
\end{array}
\right)
CZ
\left(
\begin{array}{c}
PF\cdot PFS_q(v)\\
PF\cdot PF
\end{array}
\right)\\
&=&
\left(
\begin{array}{c}
P\\
P
\end{array}
\right)
CZ^\dagger
\left(
\begin{array}{c}
FF\\
FF
\end{array}
\right)
CZ
\left(
\begin{array}{c}
FFS_q(v)\\
FF
\end{array}
\right)\\
&=&
\left(
\begin{array}{c}
P\\
P
\end{array}
\right)
CZ^\dagger
\left(
\begin{array}{c}
FFFF\\
FFFF
\end{array}
\right)
CZ
\left(
\begin{array}{c}
S_q(v)\\
I
\end{array}
\right)\\
&=&
\left(
\begin{array}{c}
P\\
P
\end{array}
\right)
\left(
\begin{array}{c}
S_q(v)\\
I
\end{array}
\right),
\end{eqnarray*}
where $P$ is a byproduct and
\begin{eqnarray*}
CZ^\dagger(X(m)\otimes I)&=&(X(m)\otimes Z(-m))CZ^\dagger,\\
CZ^\dagger(I\otimes X(m))&=&(Z(-m)\otimes X(m))CZ^\dagger.
\end{eqnarray*}

\section{Proof of Fig.~\ref{brick2}}
\begin{eqnarray*}
&&
CZ^\dagger
\left(
\begin{array}{c}
PF\cdot PF\\
PF\cdot PF
\end{array}
\right)
CZ
\left(
\begin{array}{c}
PFS_{-q}(M_m^{-1}v)\cdot X(m)F\\
PF\cdot PF
\end{array}
\right)\\
&=&
\left(
\begin{array}{c}
P\\
P
\end{array}
\right)
CZ^\dagger
\left(
\begin{array}{c}
FF\\
FF
\end{array}
\right)
CZ
\left(
\begin{array}{c}
FS_{-q}(v)F\\
FF
\end{array}
\right)\\
&=&
\left(
\begin{array}{c}
P\\
P
\end{array}
\right)
CZ^\dagger
\left(
\begin{array}{c}
FFFF\\
FFFF
\end{array}
\right)
CZ
\left(
\begin{array}{c}
S_p(v)\\
I
\end{array}
\right)\\
&=&
\left(
\begin{array}{c}
P\\
P
\end{array}
\right)
\left(
\begin{array}{c}
S_p(v)\\
I
\end{array}
\right).
\end{eqnarray*}

\section{Proof of Fig.~\ref{brick3}}
\begin{eqnarray*}
&&
CZ^\dagger
\left(
\begin{array}{c}
X_gFe^{-i(q-h+c+b)^2/2}\cdot X_hF\\
X_fF\cdot X_eF
\end{array}
\right)
CZ
\left(
\begin{array}{c}
X_dFe^{i(q-c)^2/2}\cdot X_cF\\
X_bF\cdot X_aFe^{iq^2/2}
\end{array}
\right)\\
&=&
CZ^\dagger
\left(
\begin{array}{c}
X_ge^{-i(p-h+c+b)^2/2}FX_hF\\
X_fZ_eFF
\end{array}
\right)
CZ
\left(
\begin{array}{c}
FZ_{-d}X_ce^{iq^2/2}F\\
FZ_{-b} FZ_{-a}e^{iq^2/2}
\end{array}
\right)\\
&=&
CZ^\dagger
\left(
\begin{array}{c}
X_ge^{-i(p-h+c+b)^2/2}Z_hFF\\
X_fZ_eFF
\end{array}
\right)
CZ
\left(
\begin{array}{c}
FFX_{-d}Z_{-c}e^{ip^2/2}\\
FFX_{-b} Z_{-a}e^{iq^2/2}
\end{array}
\right)\\
&=&
CZ^\dagger
\left(
\begin{array}{c}
X_ge^{-i(p-h+c+b)^2/2}Z_h\\
X_fZ_e
\end{array}
\right)
CZ
\left(
\begin{array}{c}
X_{-d}Z_{-c}e^{ip^2/2}\\
X_{-b} Z_{-a}e^{iq^2/2}
\end{array}
\right)\\
&=&
CZ^\dagger
\left(
\begin{array}{c}
X_ge^{-i(p-h+c+b)^2/2}Z_hX_{-d}Z_{-c}Z_{-b}\\
X_fZ_eZ_{-d}X_{-b}Z_{-a}
\end{array}
\right)
CZ
\left(
\begin{array}{c}
e^{ip^2/2}\\
e^{iq^2/2}
\end{array}
\right)\\
&=&
CZ^\dagger
\left(
\begin{array}{c}
X_gZ_hX_{-d}Z_{-c}Z_{-b}e^{-ip^2/2}\\
X_fZ_eZ_{-d}X_{-b}Z_{-a}
\end{array}
\right)
CZ
\left(
\begin{array}{c}
e^{ip^2/2}\\
e^{iq^2/2}
\end{array}
\right)\\
&=&
\left(
\begin{array}{c}
P\\
P
\end{array}
\right)
CZ^\dagger
\left(
\begin{array}{c}
e^{-ip^2/2}\\
I
\end{array}
\right)
CZ
\left(
\begin{array}{c}
e^{ip^2/2}\\
e^{iq^2/2}
\end{array}
\right)\\
&=&
\left(
\begin{array}{c}
P\\
P
\end{array}
\right)
CX,
\end{eqnarray*}
where
\begin{eqnarray*}
(e^{-ip^2/2}\otimes I)e^{iq\otimes q}
(e^{ip^2/2}\otimes I)
&=&
e^{i(q\otimes q)-i(p\otimes q)}\\
&=&
e^{i(q\otimes q)}e^{-i(p\otimes q)}(I\otimes e^{-iq^2/2})
\end{eqnarray*}
and we have used $e^{A+B}=e^Ae^Be^{-[A,B]/2}$ which is valid
if $[A,[A,B]]=[B,[A,B]]=0$.


\end{document}